\documentclass[a4paper,11pt]{article}
\usepackage{pos}
\usepackage{multirow}
\usepackage{lineno}

\newcommand{\snn}{\ensuremath{\sqrt{s_{_{\rm NN}}}}}
\newcommand{\pt}{\ensuremath{p_{\rm T}}}

\title{Electroweak-boson production in p--Pb and Pb--Pb collisions at the LHC with ALICE}
\ShortTitle{Electroweak bosons in p--Pb and Pb--Pb collisions with ALICE}

\author{Guillaume Taillepied\textsuperscript{$a$} for the ALICE Collaboration}

\affiliation[a]{Laboratoire de Physique de Clermont, Université Clermont-Auvergne, France}

\emailAdd{g.taillepied@cern.ch}

\abstract{Electroweak bosons are sensitive probes of the initial state of heavy-ion collisions, of which a precise knowledge is required in order to disentangle initial state effects from phenomena induced by the presence of the quark--gluon plasma (QGP). The production rate of the Z and W$^\pm$ bosons is especially sensitive to the nuclear modification of the Parton Distribution Functions (PDF), and the muonic decays (Z $\rightarrow \mu^+\mu^-$ and W$^\pm \rightarrow \mu^\pm \nu$) offer medium-blind processes carrying this information to the detector where it can be directly collected. In this contribution, new measurements of electroweak bosons in p--Pb collisions at \snn{} = 8.16 TeV and Pb--Pb collisions at \snn{} = 5.02 TeV measured by the ALICE Collaboration are reported. The data are collected at forward rapidity with the ALICE muon spectrometer and are compared to theoretical predictions with and without including nuclear modifications.}

\FullConference{%
  HardProbes2020\\
  1-6 June 2020\\
  Austin, Texas}

\begin{document}


\maketitle

\section{Introduction}

Heavy-ion collisions produced at the Large Hadron Collider (LHC) allow for the study and characterization of the quark--gluon plasma (QGP) \cite{QGPReview}. A precise knowledge of the initial state of such collisions is of utmost importance if one wants to disentangle QGP-induced phenomena from other nuclear effects. The Z and W$^\pm$ bosons, produced in the hard processes during the early stages of the collision, are sensitive probes of the initial state, and especially of the nuclear modifications of the Parton Distribution Functions (PDF). Due to their high masses, the weak bosons decay rapidly, before the typical time of creation of the QGP. The analyses presented here are based on the muonic decay of the bosons. Following the insensitivity of the muons to the strong force, the whole process is medium-blind, carrying the information from the initial state to the detector where it can be collected.

Thanks to the high energies and luminosities delivered by the LHC, weak bosons are copiously produced and their production rate can be precisely measured in heavy-ion collisions. The efficiency of the ALICE detector in such collisions, combined with the coverage at large rapidities of the muon spectrometer, allow for probing the high ($\sim 10^{-1}$ to almost unity) and low ($\sim 10^{-4}$ to $\sim 10^{-3}$) Bjorken-$x$ ranges in a region of high virtuality ($Q^2 \sim M^2_{\rm Z,W}$) where the nuclear PDF (nPDF) models are poorly constrained by other experiments \cite{nPDFs}.

\section{Analysis context and procedure}

The measurements presented here are performed using data collected with the ALICE muon spectrometer \cite{alice}. It covers the pseudorapidity interval $-4 < \eta < -2.5$ in the laboratory frame. In p--Pb collisions, following the asymmetry of the beam energies the nucleon-nucleon center-of-mass is boosted with respect to the center-of-mass in the laboratory frame by 0.465. This boost is in the direction of the proton beam, which by convention moves towards positive rapidities. The rapidity acceptance therefore becomes $2.03 < y_{cms} < 3.53$ ($-4.46 < y_{cms} < -2.96$) in the p--going (Pb--going) configuration, when the proton (Pb) beam goes in the direction of the spectrometer.

The ALICE collaboration has previously published the measurements of Z and W$^\pm$ bosons in p--Pb collisions and Z bosons in Pb--Pb collisions, both at \snn{} = 5.02 TeV \cite{pPb5tev, PbPb5tevOld}. In the latter, the analysis relied on the data sample from the 2015 data taking period only. In this report, the analyses of the Z and W bosons in p--Pb collisions at \snn{} = 8.16 TeV are presented. They are based on the 2016 data taking period, with an integrated luminosity equal to 8.47 $\pm$ 0.18 nb\textsuperscript{-1} (12.77 $\pm$ 0.25 nb\textsuperscript{-1}) for the p--going (Pb--going) period. A new measurement of the Z-boson production in Pb--Pb collisions is discussed as well, now combining the 2015 and 2018 data taking periods for an increase in luminosity from $\sim 225 \mu$b\textsuperscript{-1} to $\sim 750 \mu$b\textsuperscript{-1}. Finally the current status of the measurement of the W-boson production in Pb--Pb collisions is reported. The results on the Z-boson production in p--Pb and Pb--Pb collisions are available in \cite{paper}.

The Z-boson signal is extracted by combining muons in pairs of opposite charge. It is characterized by muons of high transverse momenta, and a dimuon invariant mass distribution centered around the Z-boson mass. The measurement is thus performed in a fiducial region defined by the acceptance of the spectrometer, a selection on the transverse momentum of the single muons (\pt{}~$>$~20~GeV/$c$) and the dimuon invariant mass (60 $< m_{\mu^+\mu^-} <$ 120 GeV/$c^2$). In this region, the nearly background-free signal is extracted by simply counting the entries in the invariant mass distribution. The extracted signal is corrected for the acceptance-times-efficiency ($A \cdot \epsilon$) of the detector, estimated by means of simulations of the process and of the detector response.

In the muonic decay of the W$^\pm$ boson, due to the presence of a neutrino, one does not have access to the full kinematic information on the final state. The signal is extracted from the inclusive single muon \pt{} distribution, which is fitted with a combination of Monte-Carlo templates accounting for the various contributions. Here as well the measurement is performed in a fiducial region defined by the acceptance of the spectrometer and a selection on the transverse momentum of the muon, \pt{}~$>$~10~GeV/$c$, and the measured yield is corrected for the $A \cdot \epsilon$ of the detector.

\section{Results}

\begin{figure}[h]
  \centering
  \includegraphics[width=0.465\linewidth]{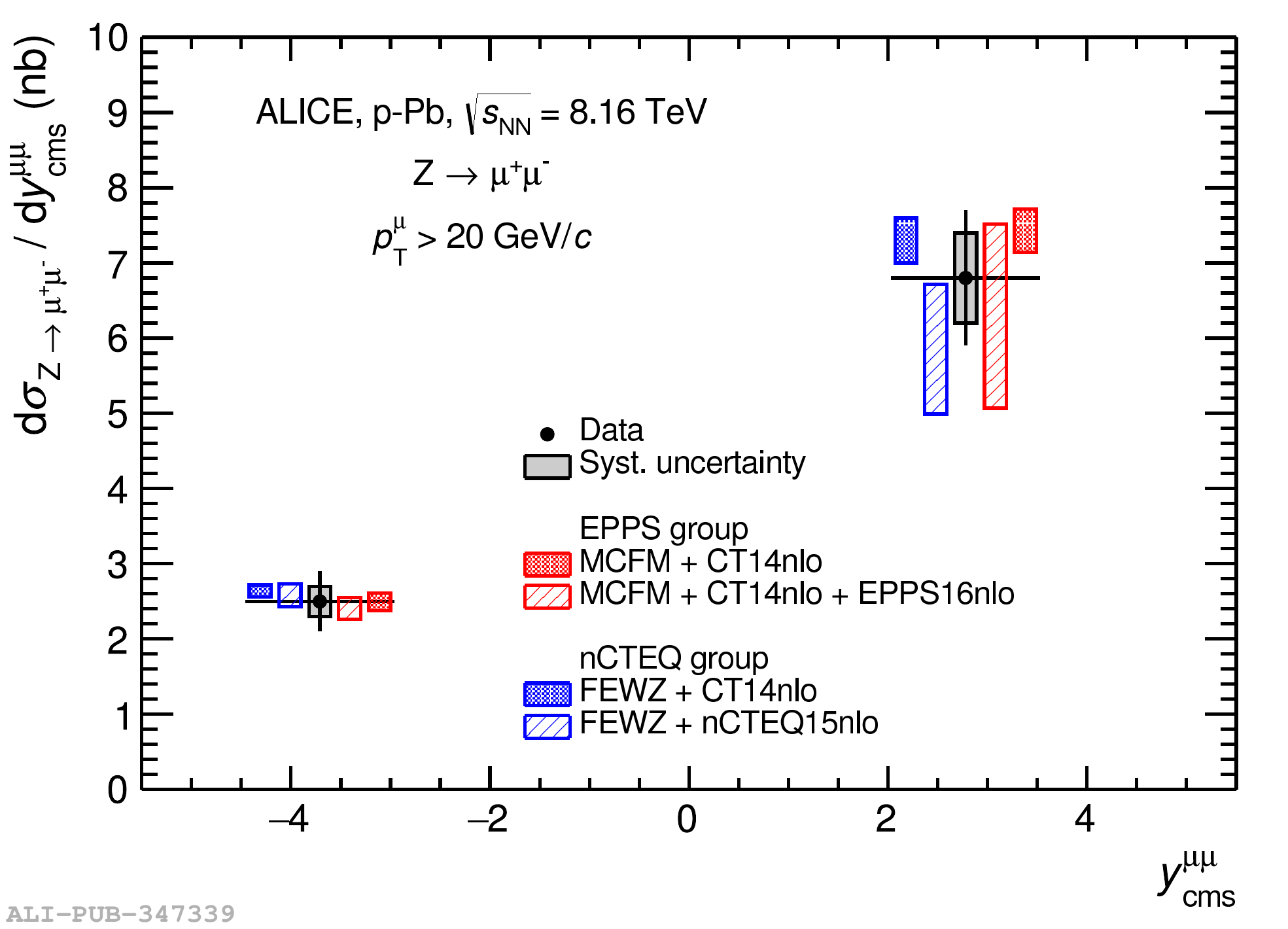}
  \includegraphics[width=0.525\linewidth]{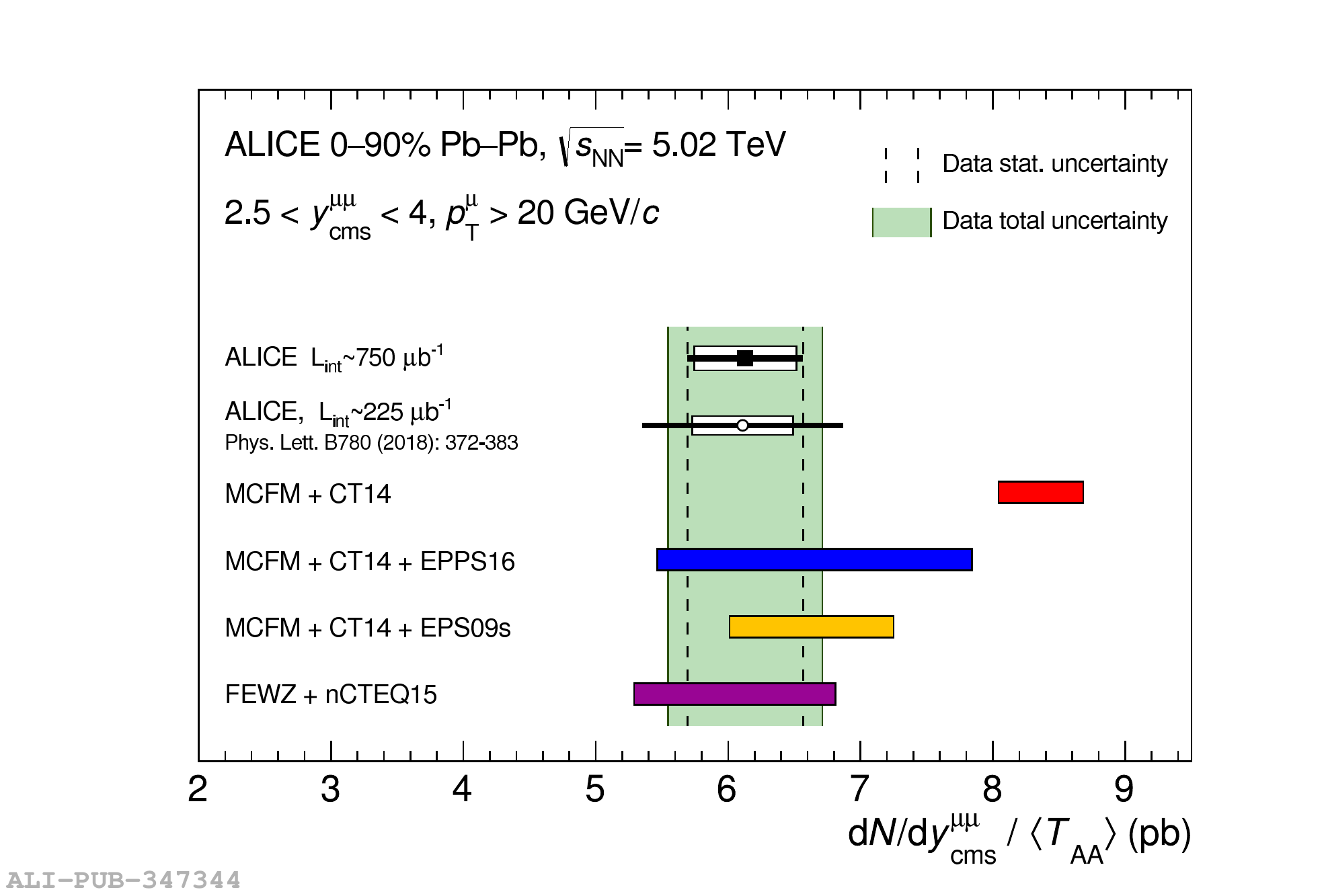}
  \caption{\textbf{Left}: Production cross section of $\mu^+\mu^-$ from Z-boson decays, measured in p--Pb collisions at \snn{} = 8.16 TeV and compared to theoretical predictions. The vertical bars and boxes around the data points indicate the statistical and systematic uncertainties, respectively. The theoretical predictions are horizontally shifted for better readability. \textbf{Right}: Invariant yield of Z $\rightarrow \mu^+\mu^-$ divided by $\left< T_{\rm AA} \right>$ measured in Pb--Pb collisions at \snn{} = 5.02 TeV. The vertical dashed band represents the statistical uncertainty on the data while the green filled band corresponds to the quadratic sum of statistical and systematic uncertainties. The result is compared with the previous ALICE measurement in the same collision system \cite{PbPb5tevOld} as well as theoretical predictions. The figures are taken from \cite{paper}.}
  \label{fig:z}
\end{figure}

The Z-boson production cross section, measured in p--Pb collisions at \snn{} = 8.16 TeV at forward and backward rapidities, is shown in the left panel of Fig. \ref{fig:z}. It is compared with theoretical predictions of the process, excluding or including nuclear modifications. In the former case the CT14 \cite{ct14} model was used, without including nuclear modifications but accounting for the isospin effect. In the latter case, nPDF predictions are obtained using either the nCTEQ15 \cite{ncteq15} nPDF model or the EPPS16 \cite{epps16} nuclear modification function combined with CT14. Although the measurements are well reproduced by the calculations, they are in agreement with predictions both including and excluding nuclear modifications, such that no firm conclusion can be drawn.

Fig. \ref{fig:z} (right) displays the Z-boson invariant yield, divided by the nuclear overlap function $\left< T_{\rm AA} \right>$, in Pb--Pb collisions at \snn{} = 5.02 TeV. By comparing the two top points, one can appreciate the increase of precision brought by the merging of the 2015 and 2018 data samples. One observes a good agreement between the data and theoretical predictions including nuclear modifications of the PDF in three different models: EPS09 \cite{eps09}, EPPS16, and nCTEQ15. On the contrary, the measurement deviates by 3.4$\sigma$ from the calculation without nuclear modification, showing the strongest evidence of nuclear modifications in all the gauge bosons analyses performed by the ALICE collaboration. The results on the Z-boson production in the two collisions system are available in \cite{paper}, where differential studies on the Z-boson production and nuclear modification factor in Pb--Pb can be found.

\begin{figure}[h]
  \centering
  \includegraphics[width=0.49\linewidth]{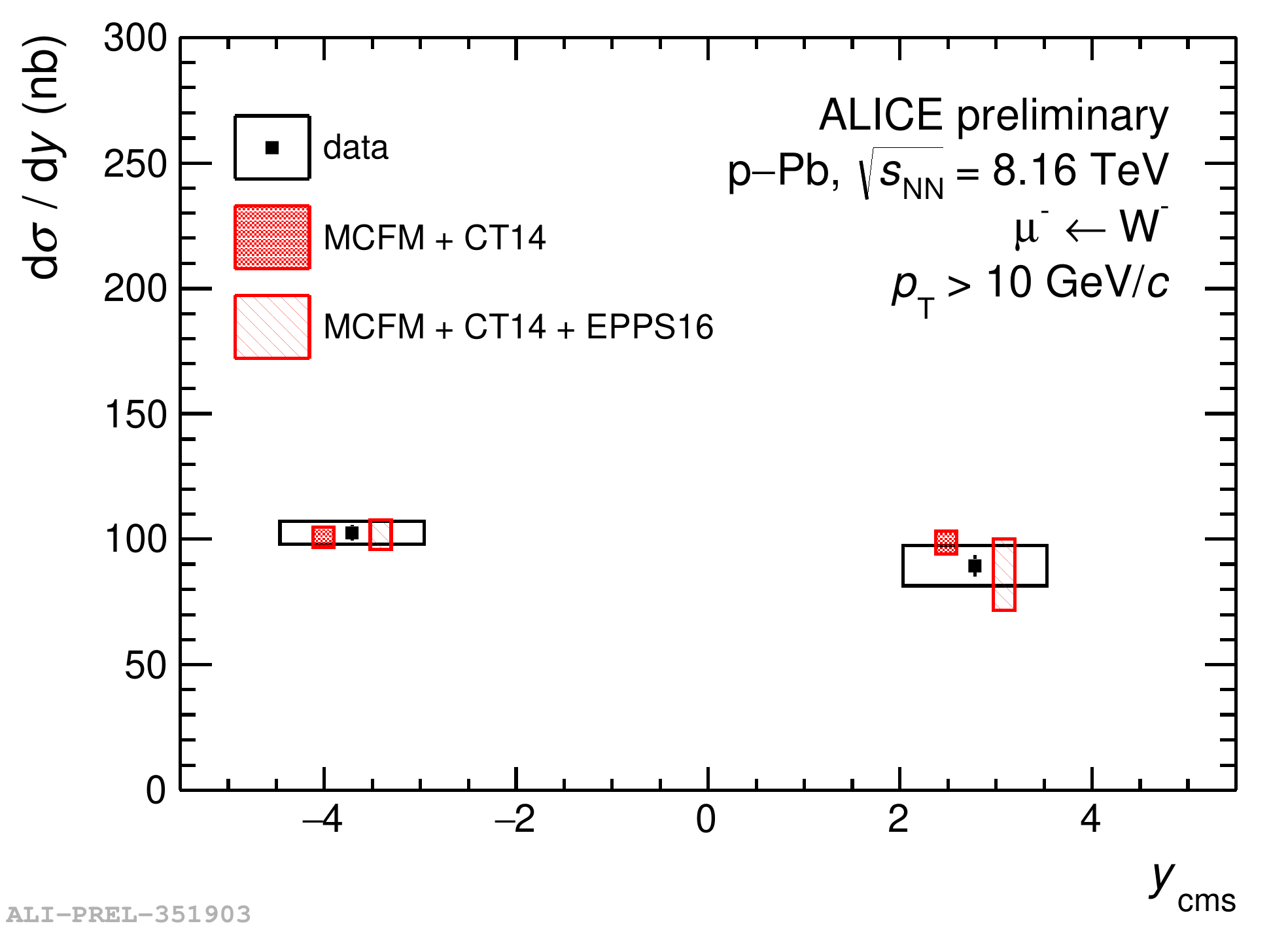}
  \includegraphics[width=0.49\linewidth]{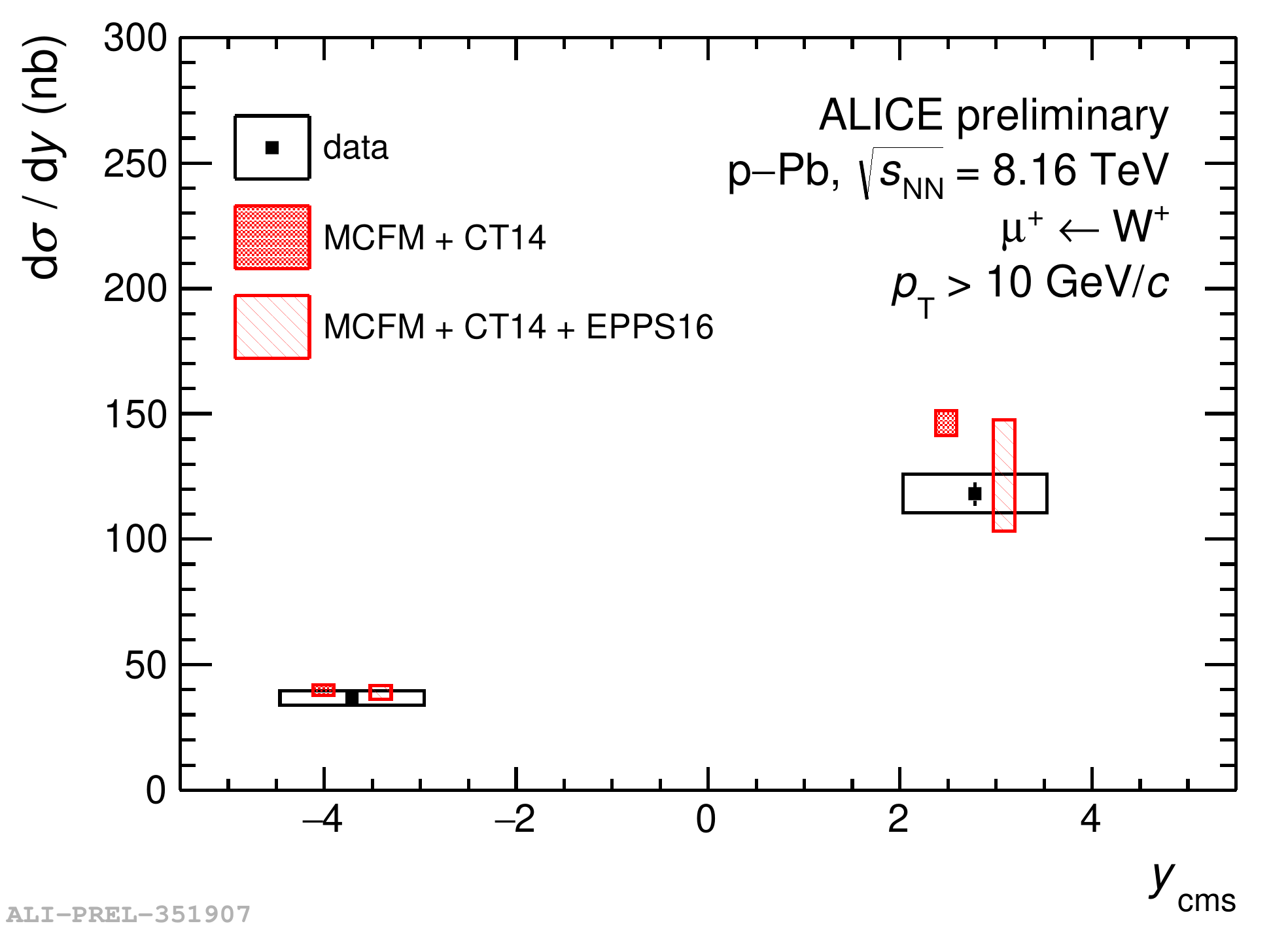}
  \caption{\textbf: Production cross section of $\mu^-$ (left) and $\mu^+$ (right) from W-boson decays, measured in p--Pb collisions at \snn{} = 8.16 TeV and compared to theoretical predictions. The vertical bars and boxes around the data points indicate the statistical and systematic uncertainties, respectively. The theoretical predictions are horizontally shifted for better readability.}
  \label{fig:WpPb}
\end{figure}

The production cross sections of W$^\pm \rightarrow \mu^\pm$ in p--Pb collisions at \snn{} = 8.16 TeV are shown in Fig. \ref{fig:WpPb} for negative (left) and positive (right) muons. They are compared with theoretical predictions based on CT14 as free-PDF set, with or without the application of the EPPS16 nuclear modification function. For the four measurements, predictions from nPDF are found to reproduce the data well. When the nuclear modifications are not applied, a significant deviation, by 2.7$\sigma$, is observed for the W$^+ \rightarrow \mu^+$ cross section at positive rapidities. This indicates some constraining power from p--Pb data on nPDF models in the low-$x$ region, where the theoretical uncertainties remain high. 

\begin{figure}[h]
  \centering
  \includegraphics[width=0.465\linewidth]{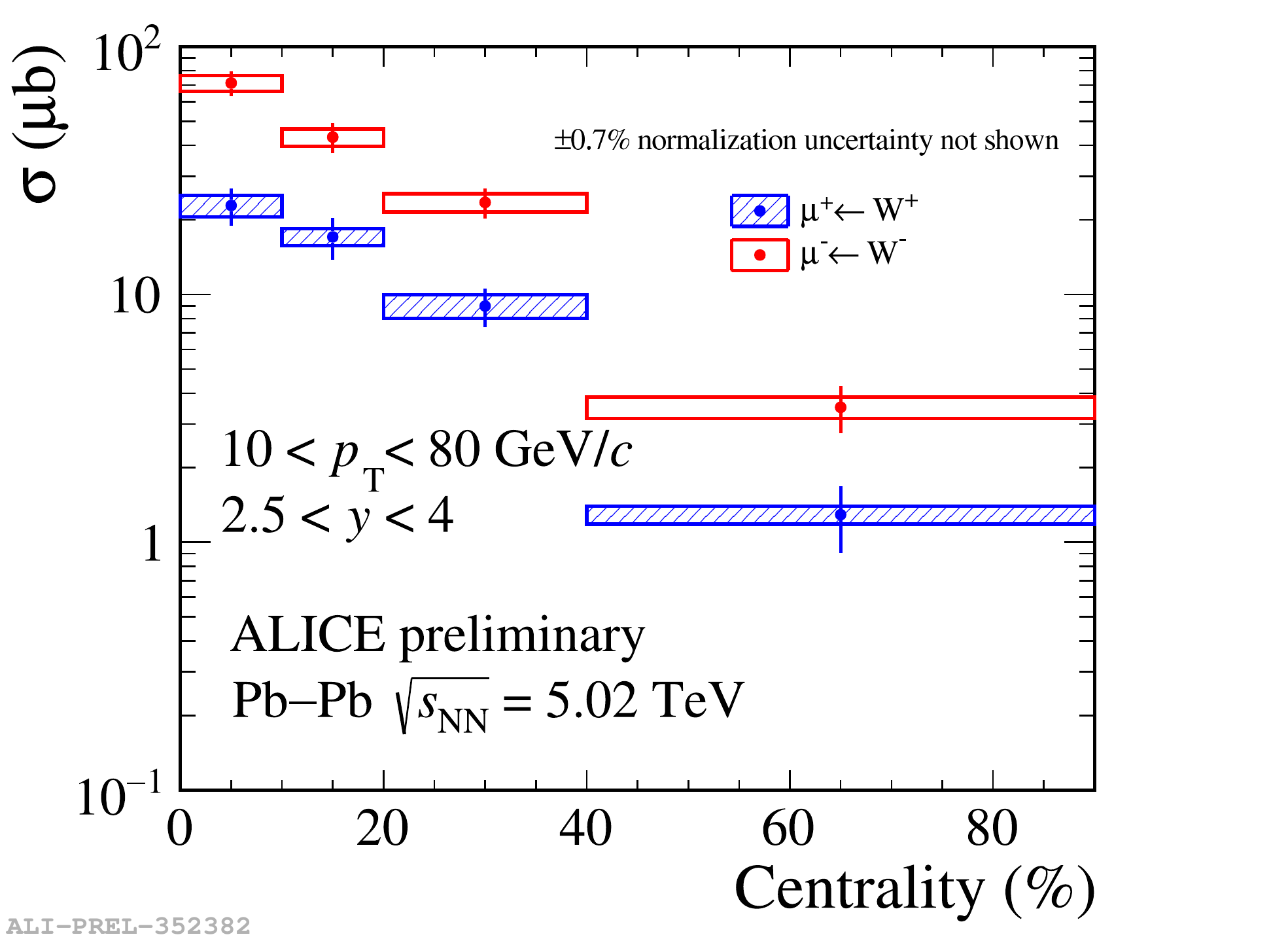}
  \includegraphics[width=0.525\linewidth]{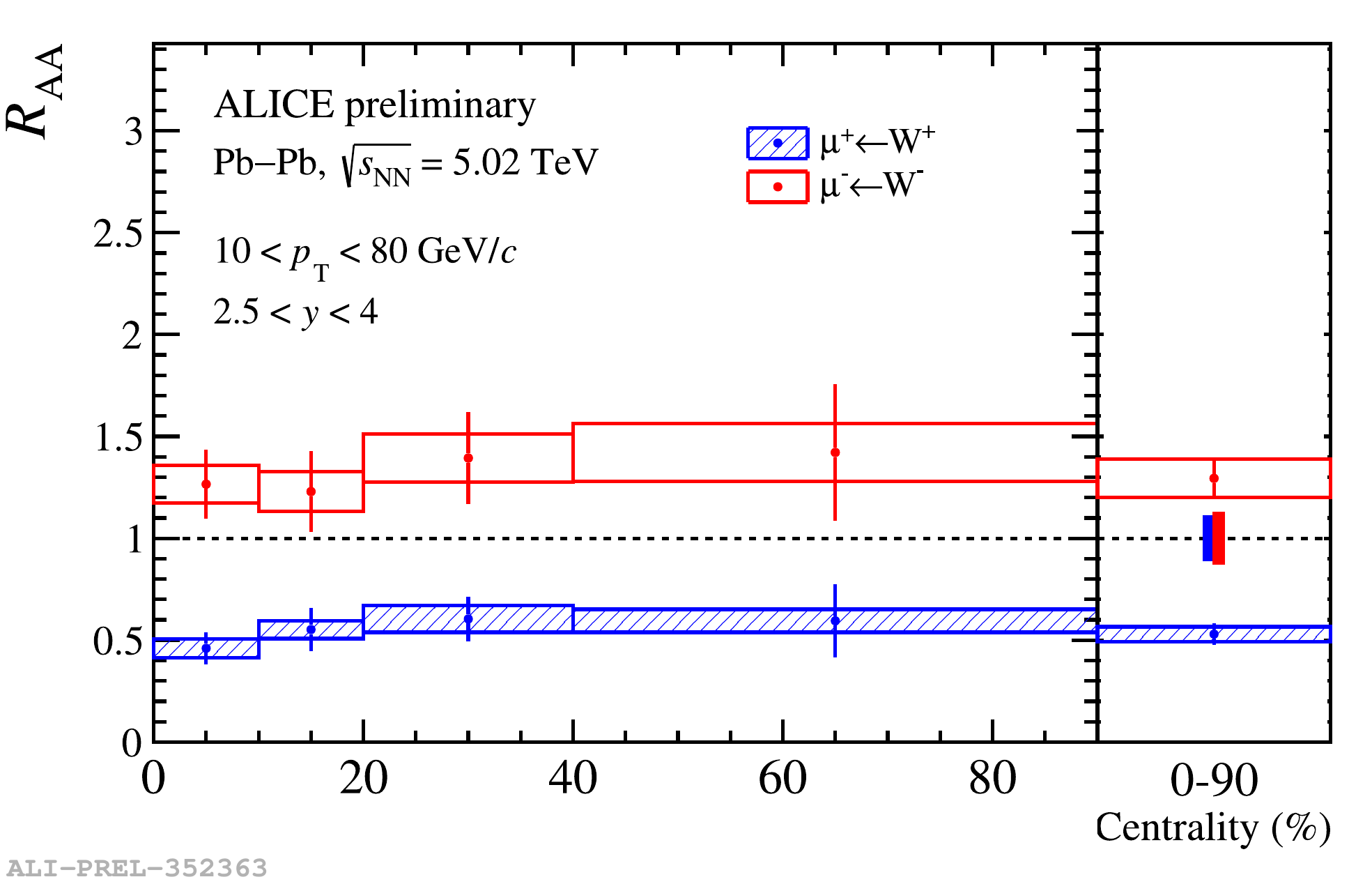}
  \caption{\textbf{Left (right)}: $W^\pm \rightarrow \mu^\pm$ production cross section (nuclear modification factor) as a function of centrality in Pb--Pb collisions at \snn{} = 5.02 TeV. The bars and boxes around the points represent the statistical and systematic uncertainties, respectively. The theoretical uncertainty on the pp reference cross section is shown as boxes around the unity line in the right panel.}
  \label{fig:WPbPb}
\end{figure}

The W$^\pm \rightarrow \mu^\pm$ production cross section in Pb--Pb collisions at \snn{} = 5.02 TeV is presented in the left panel of Fig. \ref{fig:WPbPb}. It is measured as a function of centrality, and displays the expected decrease of production when going towards more peripheral events. The nuclear modification factor $R_{\rm AA}$ is obtained by dividing the yield by the nuclear overlap function $\left< T_{\rm AA} \right>$, which creates the expected scaling versus centrality \cite{atlas} observed in Fig. \ref{fig:WPbPb} (right), and then dividing by the pp reference cross section taken from calculations with the CT10 PDF set \cite{ct10}. The $R_{\rm AA}$ shows a clear deviation from 1 due to nuclear effects, including isospin. The measurements are to be compared with theoretical predictions to investigate their ability to constrain nPDF models.

\section{Conclusion}

New measurements of the Z- and W-boson production in p--Pb and Pb--Pb collisions performed by the ALICE collaboration have been reported. They are generally well reproduced by theoretical calculations including nuclear modifications. Deviation from free-PDF predictions are observed, by 3.4$\sigma$ for the Z-boson production in Pb--Pb collisions and 2.7$\sigma$ in the W-boson analysis in p--Pb collisions. This points towards the possibility to use those measurements to help constraining nPDFs in a global fit procedure.

\end{document}